\documentclass[10pt,twoside,BCOR7mm,DIV15,headinclude,footexclude,
               cleardoubleempty,idxtotoc]
{scrartcl}

\usepackage{natbib}
\usepackage[font=small,labelfont=bf]{caption}
\usepackage[english]{babel}
\usepackage{graphicx}
\usepackage{hyperref}
\usepackage{scrpage2}
\usepackage{ifthen}
\usepackage{booktabs}
\usepackage{amsmath}
\usepackage{amssymb}
\usepackage{multicol}
\usepackage{float}
\usepackage{hyperref}

\hypersetup{breaklinks=true
,colorlinks=true,linkcolor=black,urlcolor=blue
,citecolor=black}

\addto\captionsenglish{%
}

\makeatletter
\renewcommand{\@biblabel}[1]{}
\renewcommand{\@cite}[2]{%
{#1\ifthenelse{\boolean{@tempswa}}{,#2}{}}}
\makeatother
\ofoot{\thepage}
\ifoot{}

\setcapindent{0em}
\setheadsepline{1pt}

\setkomafont{pagehead}{\normalfont\sffamily}
\setkomafont{pagenumber}{\normalfont\rmfamily}

\makeatletter
\newcommand{\listofcontributions}{\@starttoc{con}}

\newcommand{\l@contribution} {\@dottedtocline{1}{1.5em}{2.3em}}
\makeatother

\newenvironment{contribution}{
\setcounter{section}{0}
\setcounter{figure}{0}
\setcounter{table}{0}
}{
\newpage
\lehead{}
\rohead{}
}

\def\aj{AJ}%
\def\araa{ARA\&A}%
\def\apj{ApJ}%
\def\apjl{ApJ}%
\def\mnras{MNRAS}%
\def\nat{Nature}%
\def\jqsrt{J.~Quant.~Spec.~Radiat.~Transf.}%
\begin{document}

\setlength{\baselineskip}{2.5ex}

\begin{contribution}

\lehead{P.W. Morris}

\rohead{Measuring $\eta$ Carinae in the IR through sub-mm}

\begin{center}
{\LARGE \bf Measuring $\eta$ Carinae's High Mass Ejecta in the Infrared and Sub-millimeter}\\
\medskip

{\it\bf Patrick W. Morris$^1$}\\

{\it $^1$IPAC, California Institute of Technology, Pasadena, CA, USA}

\begin{abstract}
I address uncertainties on the spatial distribution and mass of the dust formed in $\eta$ Carinae's Homunculus nebula with data being combined from several space- and ground-based facilities spanning near-infrared to sub-mm wavelengths, in terms of observational constraints and modeling.  Until these aspects are better understood, the mass loss history and mechanisms responsible for $\eta$ Car's enormous eruption(s) remain poorly constrained.
\end{abstract}
\end{center}

\begin{multicols}{2}

\section{$\eta$ Car's dusty Homunculus}

The LBV and massive binary system $\eta$ Car is one of the most extensively observed objects in stellar astronomy and has the highest flux density of any object at 25 $\mu$m (second only to Jupiter and Mars during opposition), yet the nature of the Homunculus nebula's infrared emission remains poorly understood.  Prior to space-borne infrared observations, the total (gas + dust) mass of the nebula was estimated to be around 2.5 M$_\odot$, distributed in the lobes and equatorial skirt in two principle thermal components of 250 and 400 K (e.g. Davidson \& Humphreys 1997).  Infrared Space Observatory (ISO) spectroscopy covering 2.4 - 200 $\mu$m revealed much higher emission at $\lambda \; >$ 20 $\mu$m than expected from 250~K dust, and by fitting blackbody SEDs modified with an emissivity factor suited to silicate grains $F_\nu \sim \nu^{1.22} {B_\nu}(T_d)$ to the thermal continuum, \citet{M99} (hereafter M99) derived a 3-component model of $T_d$ = 110, 190, and 400 K with a total dust mass M$_{\rm{dust}} \; \geq$ 0.17 M$_\odot$, or M$_{\rm{total}} \; \geq$ 17 M$_\odot$ for an ISM-like gas to dust ratio $\rho_{gd}$ = 100.

\begin{figure}[H]
\begin{center}
\includegraphics[width=6.5cm]{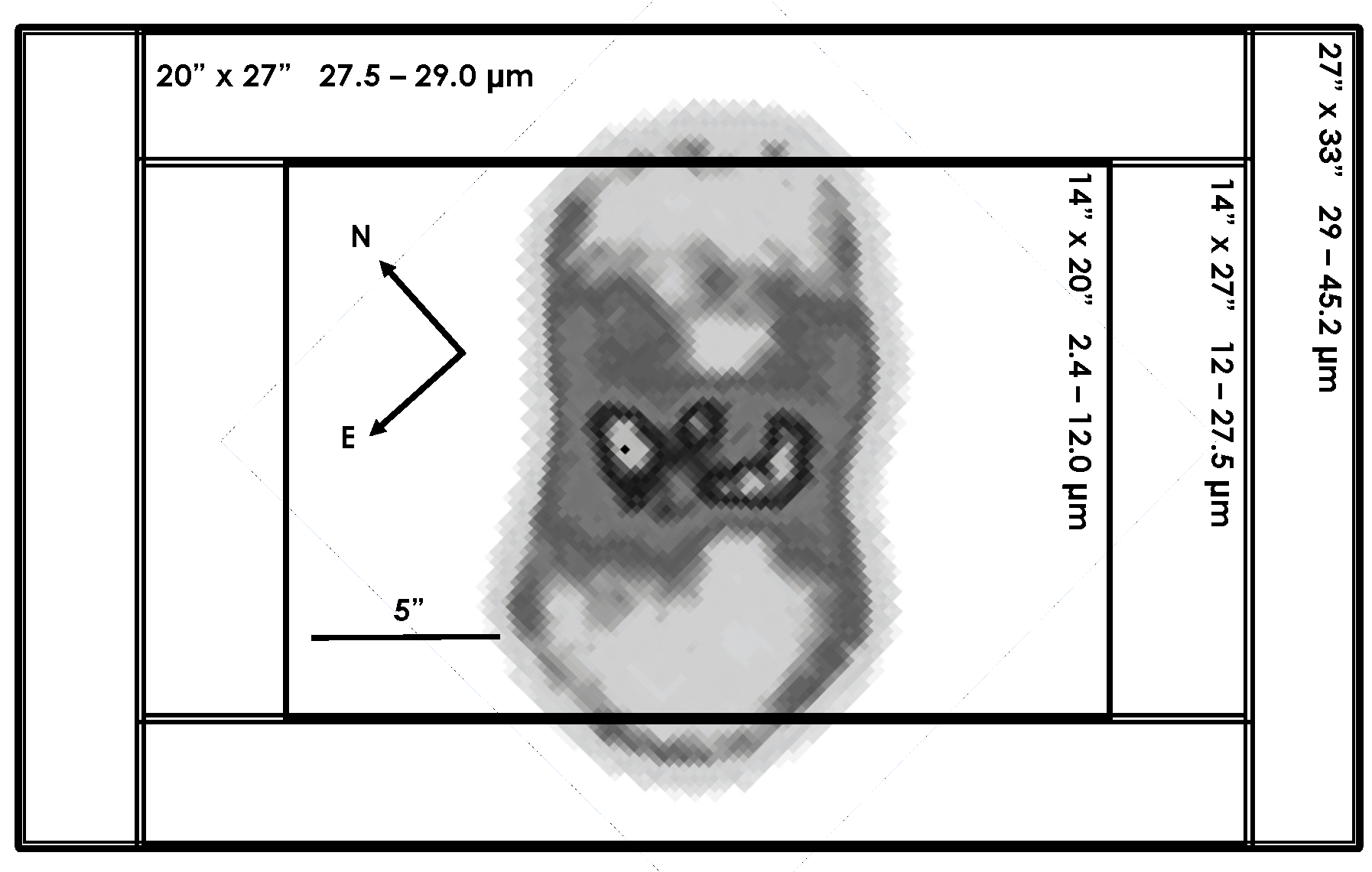}
\caption{Schematic layout of the ISO/SWS apertures on $\eta$ Car during 1996 and 1997 observations.  The 17 $\mu$m image of the Homunculus is from M99.
\label{morris:apertures}}
\end{center}
\end{figure}

The main questions about $\eta$ Car's dust are its location, especially the cool dust containing the bulk of the mass, and its composition given the uniqueness of  nebular C- and O-depletion and N-enhancement \citep{Hill01}. The distribution is key to understanding the potential role of continued formation and destruction in the dynamic central region, i.e., in the colliding winds of the erupting OB-type primary star and its putative Wolf-Rayet star companion, versus the expanding lobes where the bulk of the dust would be condensed following the 1843 eruption that formed the nebula.   The distribution could not be directly measured with ISO due to the large apertures of the SWS and LWS instruments compared to the 10$'' \times 18''$ nebula, but have been inferred by means discussed below, aided by recent observations of the sub-mm continuum and molecules such as CO, CH, HCO, and other hydrated N- and O-bearing species with the spectrometers onboard the {\it{Herschel}} Space Observatory (P.I. T. Gull; see these proceedings).   The molecules can trace physical conditions at different locations in the Homuculus that can support dust formation and survival.  An examination of the molecular gas in relation to the dust is a subject of our forthcoming paper.

\section{Dust distribution, chemistry, and mass in the Homunculus}

As mentioned, the apertures of the ISO spectrometers were large compared to the size of the Homunculus, however they did not completely encompass all nebula emission as assumed by \citet{S03} and Gomez et al. (2006).   The rectangular apertures were oriented with their cross-dispersive (spatial) minor axes parallel to the major axis of the Homunculus; see Fig.~\ref{morris:apertures}.   The progressive increasing size of the apertures in this direction with wavelength, plus the peaked beam profile shapes should create discontinuities in the SED peaking near 25 $\mu$m from an extended emitting source in this setup, as observed in many ISO spectra of nebulae such as NGC~7027 and AG Car.   The lack of beam-related jumps in the calibrated SED lead M99 to conclude that the emitting region is more compact than extended where the cool emission peaks, originating from within 6-8 arcsec of the core.  M99 suggested that the source is unresolved in the mid-IR, perhaps in a torus geometry that is unlreated to previously detected features, not that it is 1-2 arcseconds in size (as erroneously attributed by Davidson \& Smith 2000).  On the contrary Fig.~2 in M99 quite clearly points to the extended equatorial features.  Some cool dust is expected in the lobes on the basis of resolved H$_2$ observations (Steffen et al. 2014), however the observational constraints on the {\it{bulk}} of this component are inconsistent with primary location in the lobes as claimed by \citet{S03} based on an incorrect supposition about the ISO apertures and ground-based imaging at wavelegths where the cool dust is undetectable.

\begin{figure}[H]
\begin{center}
\includegraphics[width=7.0cm]{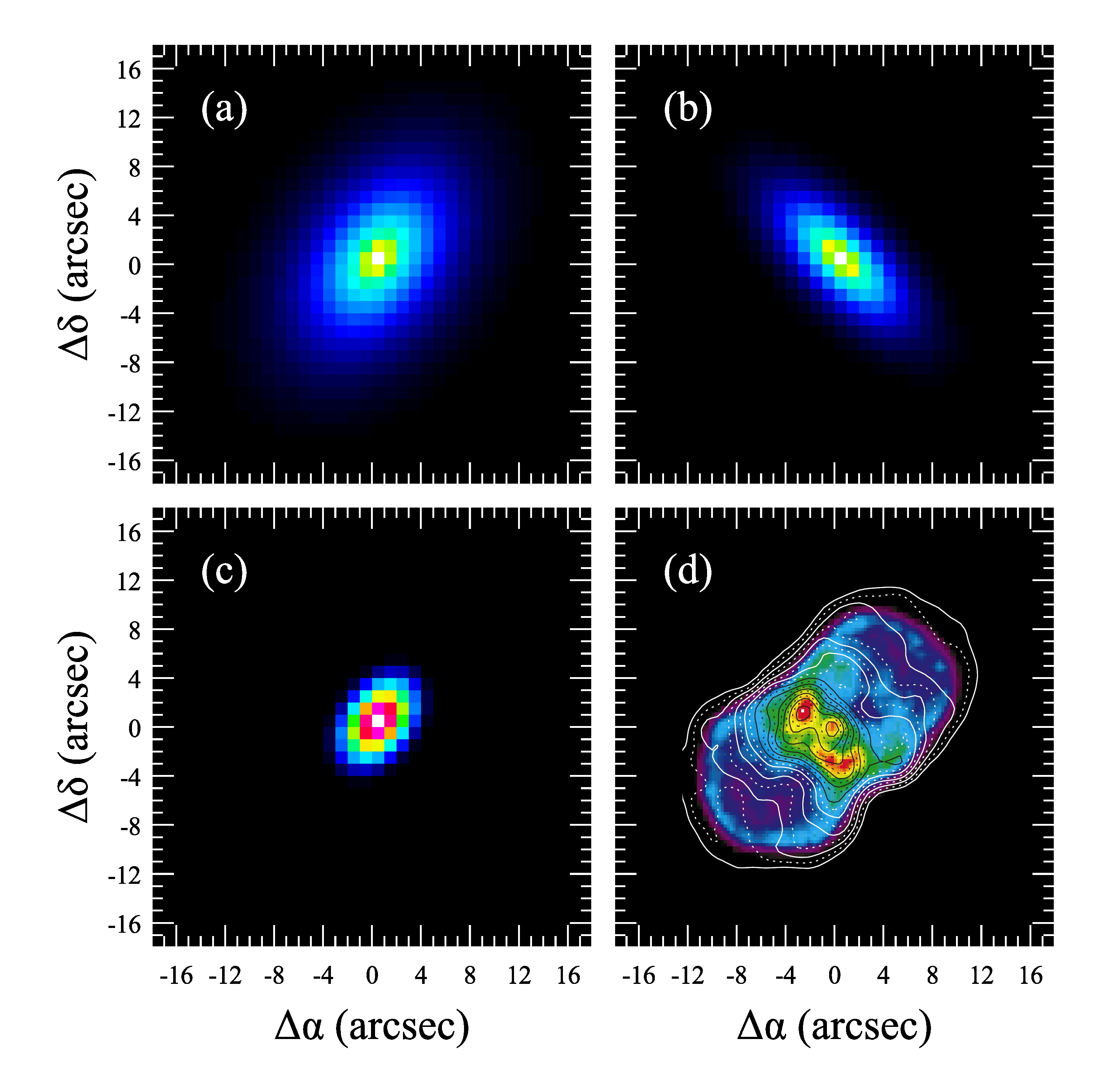}
\caption{Profiles from best-fit extended source corrections to the SPIRE spectra for elliptical profiles, scale diameters, and rotation angles of:  (a)  Sersic, 7$''$.0, 50$^\circ$; (b) Sersic, 9$''$.5, 140$^\circ$; and (c) Gaussian, 6$''$.5 (FWHM), 50$^\circ$.   A comparison TIMMI image of 17 $\mu$m emission with 10 $\mu$m contours is shown in panel (d), from M99.
\label{morris:profiles}}
\end{center}
\end{figure}
\vspace*{-6mm}

Our {\it{Herschel}} observations sampling numerous molecular lines and the cool thermal emission (T. Gull, these proceedings) support the conclusion of a dominant compact source of emission.  The SPIRE spectrometer covering 194-601 $\mu$m has variable beam sizes with a drastic difference at $\sim$305 $\mu$m where the short and long wavelength modules overlap, causing a discontinuity of the two spectral fragments in the standard pipeline processing using point source calibrations on the one extreme, or fully extended calibrations on the other.  This can be corrected interactively with different source profiles and semi-extended sizes (see Fig.~\ref{morris:profiles}), and indeed a best fit after removing background emission using off-axis detector elements is found with an elliptical sersic source profile with an effective diameter of 7$''$.0. This correction ``closes the gap'' between the SPIRE short and long wavelength spectra, and provides very good continuity (within 10\%) with the ISO spectrum.  Similar corrections are obtained in observations collected with the high velocity resolution HIFI instrument.  Such agreement between ISO and {\em{Herschel}} observations is not expected {\it{a priori}} due to possible variability in the FIR continuum between the 1996-1997 and 2011-2013 observations. These data provide only indirect indications of the emitting source size.  ALMA observations and our Cycle 3 SOFIA imaging program (P.I. P. Morris) over 19-38 $\mu$m will provide more direct measurements of the cool dust distribution.

The full SED of $\eta$ Car comprised of the ISO and {\it{Herschel}} data as well as ALMA observations obtained by \citet{A14} and other ground-based sources, spanning 2 to 7000 $\mu$m, is shown in Fig.~\ref{morris:fullsed}.  The ground-based observations are all taken with large beams except for the ALMA data which were obtained with beams 0$''$.45 to 2$''$.88 FWHM from short to long wavelengths.  With this extended dataset I have updated the 3-component model and obtain revised best-fit dust temperatures $T_d$ = 400, 170, and 95 K, and respective dust masses of 3$\times$10$^{-4}$, 0.095, and 0.17 M$_\odot$. The two coolest components have lower $T_d$ than estimated from the original ISO data alone by M99.  These values are markedly lower than deduced by \citet{S03} who attempted to re-fit the same ISO observations published by M99, but used photometrically mis-calibrated data (the standard archival data they used are uncorrected for high-flux non-linearities and detector hysteresis), and failed to recognize the solid state dust features over the 18-40 $\mu$m range, fitting through these instead and thus arriving at faulty tempertures and dust masses.

The method of fitting Planck curves to the SED provides a decent approximation of dust temperatures and masses, but for the dust chemistry and grain properties a full radiative transfer solution that models the observed solid-state features is required.  This is the subject of our forthcoming paper, meanwhile I can summarize that fitting has been carried out using laboratory optical constants for a range of amorphous and crystalline silicates, glasses, metal oxides, pure metals, sulfides, and carbon-based dust (e.g. J{\"a}ger et al. 2003), and absorption coefficients for metal nitrides \citep{P06}.  By and large the dust chemistry of the Homunculus is not surpising with a mix of Ca-, Fe-, Mg-, and Al-rich silicates (olivines, pyroxenes) and metal oxides (e.g. corundum) as well as a substantial presence of pure Fe grains.  Each has a different characteristic $T_d$, where the coolest dust appears to be dominated by the Fe dust and to a lesser degree corundum which produce broad, smooth emission bands. 

The large ($\approx$0.5-10 $\mu$m diameter) grains are amorphous and crystalline, indicating condensation conditions of both slow annealing at high gas densities, and rapid cooling at lower densities.  Whether there is a preference of either for the central region, skirt, or lobes is unknown, but it would not be unreasonable if the warm ($T_d >$ 115 K) crystalline dust which contributes less than 10\% to the total mass may be located in the lobes, possibly mixed with amorphous dust in layers exterior to the H$_2$ shells observed by \citet{S14}.  A second remarkable point is indications of nitrides, particularly in the red wing of the broad 10 $\mu$m band which is primarily composed of metal-rich silicates.   A full set of optical constants for the nitrides is not available so they cannot be included in the radiative transfer modeling, but indications of a relative contribution, if confirmed, would make $\eta$ Car's N-rich nebula the only circumstallar environment so far with Si$_x$N$_y$ dust.  Si$_3$N$_4$ dust identified as pre-solar grains in meteorites are attributed to Type II supernovae \citep{N95}, but no characteristic feature in a stellar envelope has been unambiguously identified with nitrides that could not be subsequently linked to an alternative carrier (cf. Pitman et al. 2006).

\begin{figure}[H]
\begin{center}
\includegraphics
  [width=\columnwidth]{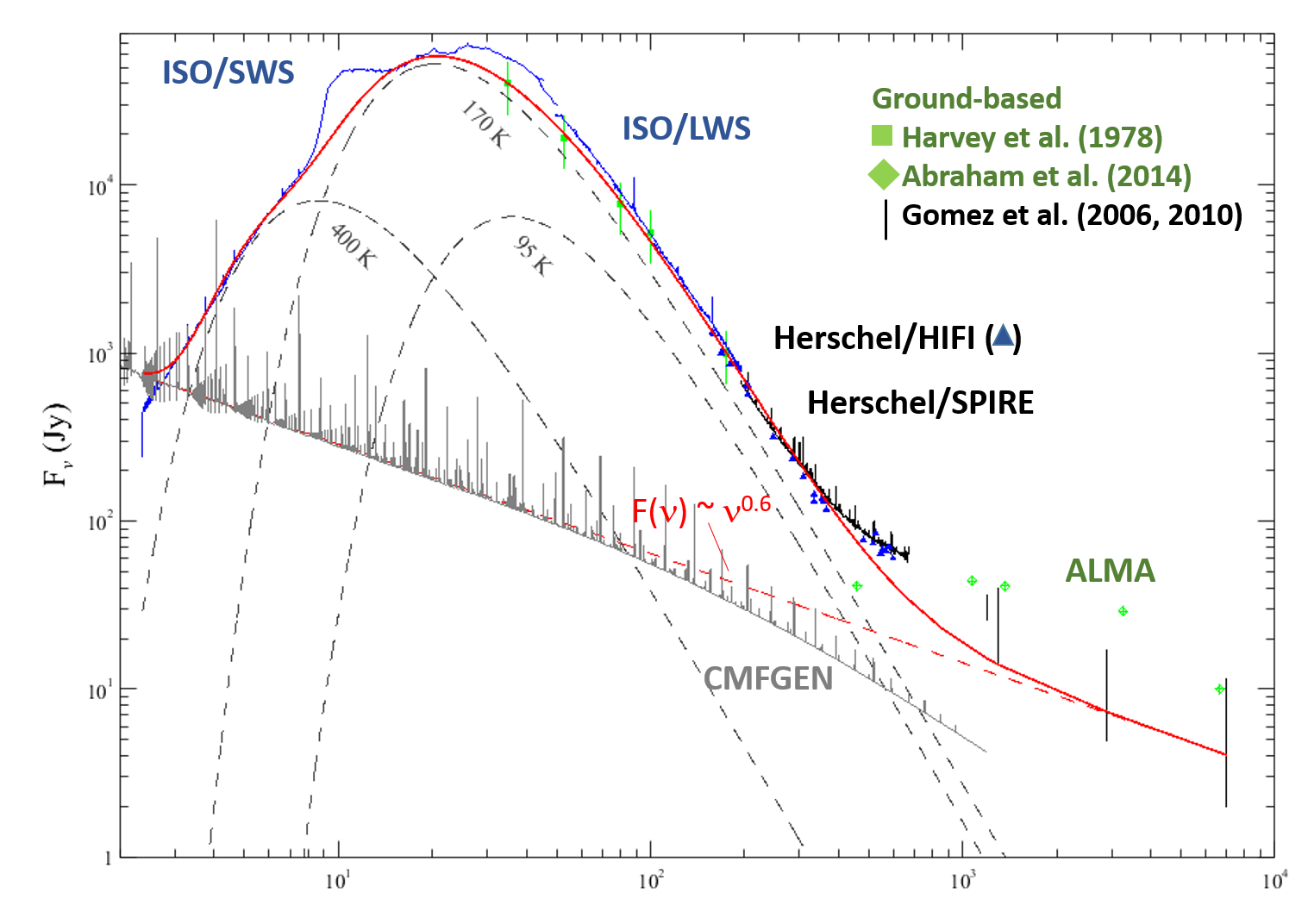}
\vspace*{-6mm}
\caption{The IR to sub-mm SED of $\eta$ Car from combined ISO, {\it{Herschel}}, and ground-based data as labeled. 
\label{morris:fullsed}}
\end{center}
\end{figure}
\vspace*{-6mm}

A total mass estimate for the Homunculus is obained by summing the individual dust contributions scaled by the associated value of $\rho_{gd}$.
A strong contribution of metallic Fe implies that $\rho_{gd}$ is much higher than the ISM value as discussed by \citet{Gail05}.  This is because only a fraction of the Fe gas condenses into grains (10-15\% into pure Fe or FeS) leaving most Fe in the gas phase which is abundantly evident in $\eta$ Car's spectrum (e.g. Hillier et al. 2001), and yielding a value of $\rho_{gd}$ much higher than 100.   Conversely $\rho_{gd}$ based on the silicate and metal oxide content could be more ISM-like, as free atomic oxygen and carbon are depleted, present in the gas phase as molecules (OH, CH, CO, HCN, etc.).  

Summing the contributions of dust at values of $\rho_{gd}$ = 200 for the Fe dust and 100 for the silicates and oxides, a {\it{lower}} limit estimate of the total mass of the Homunculus is M$_{tot} \; \geq$ 40 M$_\odot$.  This mass refers to the Homunculus, not in surrounding material as proposed by Gomez et al. (2006) who attribute excess sub-mm fluxes (see Fig.~\ref{morris:fullsed}) observed though large beams to material of a pre-1843 ouburst.  \citet{A14} have shown with ALMA that this component (also detected with {\it{Herschel}}) is associated with an unresolved source of free-free emission near the core, probably arising from wind-wind interactions of the binary system (T. Gull, priv. comm.).  

\section{Conclusions}

The distriubution of dust in $\eta$ Car's nebula, indicated to be concentrated within 6-7 arcsec around the core, is not certain without observations (e.g. ALMA, SOFIA) that can  distinguish thermal components in the central region, skirt, and bi-polar lobes.  While M99 was the first to suggest that the massive amount of dust was formed in a catastrophic close passage of the two stars, the total mass $\geq$ 40 M$_\odot$ also motivates us to consider the binary merger hypothesis as a potential scenario for the main eruption which formed the Homunculus, involving the release of kinetic energy through mass ejection from the merger of two rotating high mass stars, resulting in an eccentric orbit with a tertiary companion.   This scenario has been proposed for Be supergiants and as a general phenomenon in LBVs (Podsiadlowski et al. 2006; Suzuki et al.  2007; Justham et al. 2014).  If the bulk of the dust is concentrated in the core region, then continued formation (and destruction) in colliding stellar winds similar to dusty WC binaries must also be taken into account in the mass budget.

\bibliographystyle{aa} 

\end{multicols}

\end{contribution}


\end{document}